# Research on the evolution of smart meter technology

Zhen Zhang

Huaneng Jinan Huangtai Power Generation Co., LTD., Jinan, Shandong province 250100

abstract：smart meter is not only a device used to measure the amount of electricity, but also a core component of the smart grid, realizing the efficient monitoring, prediction and management of power use. With an insight into the evolution of smart meter technology, I realized that this change didn't happen overnight. It has undergone a long journey from the initial mechanical electricity meters to the electronic electricity meters, and now to the highly intelligent electricity meters. Technological breakthroughs at each stage have laid the foundation for the final form of smart meters. In the era of mechanical watt-hour meters, the measurement of electric energy mainly depends on the rotation and counting of mechanical structures. The accuracy and stability of this method are affected by mechanical wear, environmental interference and other factors, and it is difficult to meet the increasing demand for power management. With the rapid development of electronic technology, the electronic electricity meter came into being. It uses electronic technology to sample and process the current and voltage, which greatly improves the accuracy and stability of the measurement. At the same time, the electronic electricity meter also has the function of remote meter reading and data processing, which has brought great convenience to the power management. However, there still have some limitations, such as the complexity of data processing and the limitation of communication capacity. It is these challenges that drive the creation of smart power meters. smart meters combine advanced technologies such as the Internet of Things, big data and cloud computing to realize the real-time monitoring, analysis and prediction of the use of power. It can not only accurately measure the amount of electricity, but also provide personalized suggestions and energy saving solutions according to the user's electricity behavior and demand. In the evolution process of the smart meter technology, the communication technology plays a vital role. From the initial wired communication to the integration of wireless public network, wireless private network, power line carrier and other communication modes, the intelligent electricity meter has realized the real-time communication and data sharing with the power grid management system. This makes the power management more intelligent and efficient, and provides users with more convenient and personalized electricity experience. However, the development of smart meter technology also faces some challenges. How to ensure the security and privacy protection of data is an urgent problem to be solved. At the same time, with the continuous development of smart grid, smart meters need to be more closely connected and coordinated with other devices to achieve more efficient and reliable power management. Looking into the future, I believe that the smart meter technology will continue to innovate and breakthrough. With the further development of the Internet of Things, big data, artificial intelligence and other technologies, intelligent electricity meters will have more powerful functions and performance, and provide a more solid foundation for the intelligence and efficiency of the power industry. At the same time, we also need to pay attention to data security, technical standards, cost and other challenges, and actively seek solutions to promote the sustainable development and application and promotion of smart meter technology.

keyword：smart meter　technology evolution　power grid

Chinese chart classification number：TM933.4

about the author：Zhen Zhang　（1977-）, engaged in research on　metering technology　721047546@qq.com

## 0 Preface

In April 2009, Hangzhou Haixing Company took the lead in delivering the international report of "Application and Development Trend of smart meter" at the domestic electricity meter industry conference.

In May 2009, State Grid Corporation released the development plan of "Strong Smart Grid" (2009--2020) at the International Conference on UHV Technology held in Beijing.

Under the background of these international and domestic metrology development, in August 2009, State Grid issued China's first smart meter series enterprise standard. Since then, State Grid has carried out its 11-year research and development, large-scale application and construction of electricity information collection system on the user side.

- -State Grid has developed a wide collection of styles for smart meters

From 2009 to 2020, according to the release of smart meter standard, state Grid launched two generations of four models of smart meters:

· 2009 version of State Grid standard smart meter —— first generation smart meter (first);

· 2013 version of State Grid standard smart meter —— The first generation smart meter (paragraph 2);

· According to the 2016 State Grid design scheme: "dual-core" smart meter based on IR 46 concept —— New generation smart meter (the first);

· According to the technical requirements of State Grid in 2019: a new generation of smart meter with multi-core modular structure (paragraph 2).

- -Total amount of smart meters operated by state Grid

From 2009 to 2020, state Grid mainly installed and applied the 2009 versions and 2013 versions of state Grid standard smart meters. By the end of 2018, the total number of smart meters in operation in the state Grid power supply business area was 458 million, covering 99.87% of users, with a total investment of 99.8 billion yuan. Among them, the 2013 standard smart meters are about 331 million, accounting for 72.2% of the total smart meters in operation; the 2009 standard smart meters are about 127 million smart meters, accounting for 27.7%.

- -The accumulative total bidding amount of state Grid smart meters

Since 2009, the bidding product of state Grid smart meters is the 2009 version of standard smart meters, but after mid-2013, it was changed to the 2013 version of standard smart meters. From 2009 to 2019, the total bidding volume of smart meters was calculated from the total amount of smart meters running at the end of 2018, the inventory of smart meters and the metering failure processing table at the end of 2018 (about 20%) and the bidding volume in 2019. By the end of 2019, the total bidding amount of state grid smart meters was estimated at 621 million smart meters, with a total investment of 129.1 billion yuan.

- -State Grid's goal of large-scale development and application of smart meters

Given that the electricity meter is a legal measuring instrument for the settlement of electricity consumption and the assessment of the technical and economic indicators of the power grid, the state Grid expects to be in the power grid system:

· For the first time, realize the unification of the type, function, technology and testing (including meter software testing), and strengthen the management of electric energy metering technology.

· For the first time, centralized and unified automatic verification of smart meters (10 million-40 million) by the provincial power grid power Research Institute, so as to improve the quality level of smart meters in the whole network.

· In August 2018, State Grid carried out large-scale sampling inspection of smart meters (about 360,000 meters) in 26 provincial power grids for more than 7 years, and the qualified rate of sampling inspection reached 99.64%, which showed that the quality and quality deviation of smart meters in online operation were ideal. However, the statistical method of the sampling pass rate is questioned online: usually, the failure replacement rate in the initial stage of smart meters installation and operation is very high, especially the 2009 version of the state grid standard smart meters.

The above paper summarizes the development of state Grid smart meters (2009- -2020) from the development style, total amount of operation, total amount of bidding and the development and application goals of smart meters.

- -Next, review the whole development of smart meters of State Grid, and mainly conduct the application research on each type of smart meters.

· In February 2009, the Marketing Department of State Grid held an expert seminar on "Power meter Type, Function and Bidding Technical Specifications" in China Electricity Research Institute. The author of this article was invited to attend the meeting. Therefore, the author of this paper pays attention to how the "electricity meter type, function and bidding technical specification" formulated by China Grid in 2008 is transformed into the 2009 version of the state grid smart meter series enterprise standard.

· Conduct application research on various types of smart meters of State Grid, mainly extract the design characteristics of smart meters through the collection and interpretation of smart meter standards, professional journals and smart meter application reports; discuss the controversial issues related to the design and application of smart meters, and put forward suggestions for the improvement of smart meter design and application. From 2010 to 2020, the author of this article has followed the whole development of state Grid smart meters, and has written 17 special manuscripts, which have been compiled in the Collection of Modern Electricity Meter Application Research.

**1. Design characteristics of each type of smart meters of State Grid and evaluation of the quality of metering products.**

**1) The original 2009 version of state Grid standard smart meters**

　-Design features

· The 2009 version of the standard gives the definition of smart meters for the first time. Because of the "intelligence" function is not in place, causing different views;

· In terms of meter function, compared with DL / T614- - -2007 multi-function Electricity Meter, the new standard proposes the application of hard clock circuit, programming password, cost control, load record, step electricity price, preliminary application of power outage meter reading, ESAM module security certification, etc.

- -Quality evaluation: The 2009 edition of the standard is converted and supplemented by the "electricity meter type, function and bidding technical specification" formulated by the State Grid in 2008. Due to the short conversion time, lack of scientific and detailed demonstration and testing for the newly proposed function and technical requirements, and incomplete and unclear standard content, many original requirements of design and manufacturing quality of 2009 standard smart meters, mainly the incomplete and solidified power metering functions cannot be selected by other professional and provincial power grid; irregular meter operation and operation; some meter suppliers lack the design and manufacturing experience of smart meters, especially defects in software design, component screening and production process, resulting in high failure rate in the initial stage of meter operation. To this end, in July 2012, the Marketing Department of the State Grid issued the Notice on Further Strengthening the Quality Control of Smart Power Meters.

— situation of application

· From the end of 2009 to the middle of 2013, the total amount of 2009 version of State Grid was bid for 225 million smart meters, with an investment of about 46.8 billion yuan. At present, there are about 127 million standard smart meters operating online, accounting for 27.7 percent of the total smart meters in operation.

· In the middle of 2013 and later, the 2009 version of standard smart meters will no longer participate in the bidding of state Grid smart meters.

**2) 2013 version of state Grid standard smart meter**

　- -Design and test characteristics

· Compared with the 2009 version of the standard, the measurement function of the 2013 version of the standard is not complete and solidified, and cannot be selected according to demand, which is the defect of the new standard;

· The new standard puts forward many innovative and practical requirements for the metering performance and function expansion of smart meters:

It mainly selects advanced metering chips, refine event records, ensure measurement safety and prevent AC-DC power theft, formulate technical specifications for 15 categories of major components; implements centralized and unified automatic verification of smart meters in the province; and puts forward gradual configuration of HPLC and gateway requirements.

- -However, the 2013 version of the standard smart meter does not meet the requirements of the application of the (OIML) IR 46 standard in China.

— situation of application:

· In the middle of 2013- -In 2019, state Grid invited 400 million copies of the 2013 standard smart meters for bidding, with an investment of about 83.2 billion yuan. At present, there are about 331 million standard smart meters in the 2013 version running online, accounting for 72.7% of the total smart meters in operation. These 2013 version of the standard smart meters were originally scheduled to be replaced in 2019 with the "dual-core" smart meters —— new generation smart meters (the first).

· However, due to changes in the development of state Grid from January 2019- -January 2020, the 2013 version of standard smart meters continues to operate on the grid.

· On April 19,2020, online report: information of the total amount of the first batch of electricity meters in 2020 (24.752 million) and the total amount (4.26 billion yuan). It is estimated that the state grid bidding metering products, or choose the 2013 version of the state grid standard smart meters.

**3) According to the 2016 State Grid design scheme: "dual-core" smart meter based on IR 46 concept —— New generation smart meter (the first)**

State Grid "" dual-core " smart meter Design Scheme based on IR 46 concept is the design outline of new smart meter products in which OIML's IR 46 standard is first applied in China.

-Design features

· In order to implement the metering characteristic protection requirements of IR 46 standard, the architecture design of "dual-core" smart meter adopts the "dual-core" structure, designs the separation of metering core and management core, and changes the integrated architecture design scheme of smart meter in the early stage.

· "Specific design technology of dual-core" smart meter: adopt dual MCU and dual memory scheme; propose voltage and current sampling measurement or instantaneous value measurement; electrical energy data with time scale; real clock power design more reliable, adopt HPLC communication mode; and management core software for online upgrade.

- -Design scheme review

The meter name of the "dual-core" smart meter design scheme is not appropriate, the design content is incomplete, and there are many missing items, mainly including:

· "Double-core" as a part of the meter name, is not appropriate."Dual core", can only state that the meter is mainly used for the implementation of IR 46 standard. In fact, the "part of IR 46 standard 1: measurement and technical requirements", the core content is accuracy requirements, measurement characteristics protection two aspects, and IR 46 standard put forward the measurement accuracy requirements of the new system is more important.

· The body of the design scheme includes six parts: phenome design, structure design, interface definition, metering core function requirements, functional requirements of management core, and dual-core smart meter test; appendix, including communication protocol and management core program upgrade. The accuracy requirements, software separation requirements and part 2 of measurement control and performance tests.

· Missing items in traditional smart meter design, including meter function positioning and overall architecture design; hardware and software design and test; meter reliability design and verification test; meter process design and test; meter type test, the design scheme is not particularly specified.

- -The survival period of the project is only 4 years

The process of "dual-core" smart meter from project approval to prototype trial operation:

· Project approval: In January 2015, the Marketing Department of State Grid [2015] No.53 proposed: "To carry out the international application research of IR 46 active power meter".

· Design scheme: In September 2016, State Grid issued the "Dual-core" Intelligent Power Meter Design Scheme based on IR 46 concept, and organized the publicity and implementation meeting for electricity meter enterprises.

· Prototype launch: In 2017, some domestic electricity meter enterprises have developed "dual-core" smart meters, and sent them to the State Grid Measurement Center for inspection.

· Trial operation of prototype: In November 2018, the "dual-core" smart meter of State Grid was operated for the first time in Zhejiang power grid.

· Transfer to transition products: In January 2019, State Grid proposed alternative meter positioning and expanded function design requirements for "dual-core" smart meters. "Dual-core" smart meters, as the technology reserve of the new generation of smart meters, provided the design experience of the implementation of IR 46 standard for the new generation of smart meters (the second paragraph).

4) Design according to the technical requirements of State Grid in 2019: a new generation of smart meter with multi-core modular structure (paragraph 2)

In June 2019, State Grid launched a prototype of a new generation of smart meters (the second type), whose performance and functions cover not only the requirements of the implementation of IR 46 standard, but also the requirements of alternative meter positioning and expansion function design proposed by State Grid in January 2019.

-Design features

The design of the new generation smart meter (the second paragraph) is different from the previous smart meter, and adopts multi-core modular structure design.

· Multi-core design: metering core and management core separation design; the electric energy metering type extends from active and reactive measurement of sine wave to base wave active and reactive active measurement and harmonic active measurement; active energy collection and freezing period is increased from 15 minutes to 1 minute; meter software test requirements, 10 legal related software and 7 non-legal related software; the service life of the meter is increased from 8 years to 15 years.

· modular design: including measuring core module, management core module, expanded functional module, and connected into a metering system. Three typical application scenarios for the expansion module are proposed, namely, orderly charging of electric vehicles, intelligent analysis and application of residential power load characteristics, and information collection and application of "multi-meter integration".

- -Follow-up development: In this paper, the author proposes the application topics that need to be studied after the launch of the prototype, including: the formulation of a series of product standards covering all the performance and functional design of the new generation of three-phase smart meters, and the application and development projects of new technologies, see parts 3 and 1) of this paper.

- -Changes in the application situation

· In January 2020, State Grid adjusted the development direction, and the application trend of the new generation of smart meter (the second type), which attracted attention.

· Here, the author of this paper believes that the new generation of smart electric meter (the second paragraph) of State Grid adopts the multi-core module architecture design, which subverts the traditional design scheme of the integrated architecture of the electronic electric meter / early smart electric meter, and realizes a new leap in smart electric meter design. Among them, in China, the extended function module design is proposed for the first time, which can meet the requirements of the meter expanded function on demand, which is desirable. However, the product design problem of state Grid's new generation of smart meter (the second model) lies in the expansion of functional module. It has been pointed out above: in January 2019, State Grid proposed and promoted the implementation of the alternative meter positioning and expansion function design requirements, including three application scenario functions. Therefore, according to the new development direction of state Grid, the positioning and expansion function of the new generation of smart meters (the second paragraph) need to be re-approved and selected, and conclude whether they are desirable or undesirable after improvement.

- -Before state Grid's new generation of smart meters are officially launched, there is new demand

In April 2020, the online report, "State Grid Marketing Department issued the professional work points of intelligent electricity consumption in 2020" pointed out: in " actively promote the optimization of energy use in the public transformer area, promote the research and development of household intelligent appliances, combined with the new generation of smart meters, joint development of intelligent Internet air conditioning, electric water heater, etc. Build a state grid household smart energy use platform, realize the interaction and intelligent control of home appliances and the power grid, improve household energy efficiency, and reduce electricity costs."

Explain the above professional work points of state grid intelligent electricity consumption: promote the energy optimization project of the public transformer platform area, carry out the research and development of intelligent Internet home appliances through the new generation of smart meter, collect and transmit the information of the household intelligent energy platform, and carry out the interaction and intelligent control between home appliances and the power grid. These new functions can be realized in the new generation of smart meter (second) expansion function module only by adding the metering module, connecting, sensing and control module with indoor home appliances, and using the new indoor communication technology module.

## 2. Comprehensive evaluation of the development process of state Grid smart meters from 2009 to 2020

Below, by summarizing the characteristics and quality evaluation of the above smart meters of the state grid, the comprehensive evaluation of the development process of smart meters in the overseas network is refined.

### 1) Important achievements of the development of state Grid smart meters in the past 11 years

- -For the first time, basically full coverage of smart meters has been achieved in the state grid power supply business area. As mentioned in the preface of this article, by the end of 2018, a total of 458 million smart meters were in operation, and the user coverage rate reached 99.87%. Therefore, 2018 became the second turning point in the development history of electricity meters in China, that is, the main body of electricity meters in power grid operation changed from induction meter + electronic meter to the only smart meter, and induction meters all faded out of the domestic electricity meter market.

(Note: The first turning point in the history of electricity meters in China: in 2005, the total output of electronic meters exceeded induction meters for the first time)

- -For the first time to realize the unification of the quality management of smart meters in the state grid system, and accumulate the experience of the quality management of the meter at both ends and among the supervision.among, Unity of product standards: State Grid has issued 2009 and 2013 editions of smart meter standards, Realize the unification of the type, function, technology and test (including the meter software test) of the smart meter; Unification of manufacturing quality supervision: State Grid issued documents on strengthening the quality control of smart meters, Using advanced power metering chips, Technical specifications for 15 major components, Lift the meter high temperature aging process; Unified meter test: the provincial smart meters (10 million- -40 million) are centralized, unified and automated verification by the provincial power grid power research institute, Greatly improve the quality and consistency of smart meters.

- -The multi-functional design of smart meter conforms to the demand of electricity price reform and greatly expands the application scenarios of the meter.

Smart meters in DL / T614 multi-function electricity meter, has put forward new functions are: cost control, ladder price, active power collection, freezing cycle, 1 minute, collecting power with time scale, hard clock circuit and calibration, programming password application, blackout meter reading, prevent ac / dc strong magnetic field theft measures, management core software upgrade, voltage and current waveform data output, low voltage grid power failure report, ESAM module security certification, etc.

　　- -Overall architecture design of smart meter: the integrated architecture design of traditional electronic meter / early smart meter is transformed to the integrated architecture design of multi-core module of the new generation of smart meter (the second model). Among them, the separation design of metering core module and management core module meets the requirements of IR 46 standard; the implementation design of extended function module can meet the requirements of changing the function, practical, this design is very characteristic.

　　- -In terms of the reliability of operating electricity meters, the 2013 edition of State Grid standard smart meters adopts many innovative application function designs on the basis of summarizing the design defects of the 2009 edition improvement of standard smart meters. After 7 years of operation assessment, the product quality is relatively stable and relatively reliable operation. As described above, since mid-2013,331 million standard smart meters have been installed in the 2013 version, accounting for 72.2% of the total smart meters in operation. It is estimated that in 2020 and the following 2-3 years, the main metering products of state Grid smart meter bidding will be the 2013 version of state Grid standard smart meters.

**2) In 2009- -2020, what are the problems in the development and application of four smart meters of two generations of State Grid?**

　　- - smart meteris developed frequently and is eager to achieve success. It mainly considers the demand for function expansion, low price, and the large investment required for the replacement of operation meter

　　· Electronic electricity meter: according to the national legal metrological verification period is 8 years. It can be interpreted as that the operation life of the new smart meter is longer than 8 years from the perspective of ensuring qualified measurement error.

　　· Case 1:2009 edition standard smart meter project. It has been pointed out above that due to the high failure rate and irregular functional application of this smart meter, it will no longer participate in the bidding of state Grid smart meter in the mid of 2013 and later, and the shortest online operation period is only 3.5 years. In the first half of 2013, the maximum number of the smart meters operating online was 225 million, with a total investment of 46.8 billion yuan. At present, the number of online operations has dropped to 127 million, and it is estimated that in another three years, all the online operations can be replaced and eliminated.

　　· Case 2: The "Dual-core" smart meter project. The 2013 version of smart meter has only been in operation for 3 years. In 2016, State Grid was eager to release the "dual-core" smart meter Design Scheme based on IR 46 concept, and laid out the electricity meter enterprises to develop "dual-core" smart meter, which is the need of state Grid to apply OIML's IR 46 standard in China first. In fact, the author thinks: at present, the IR 46 standard in China application has limitations, mainly the IR 46 standard is only the international suggestion of active power meter, reactive power measurement and other types of meter, OIML will be launched, there is no determined period; IEC standard: it is understood, IR 46 standard IEC active power meter standard is still in the process of formulation; state grid enterprise standard: the IR 46 standard (including accuracy requirements, measurement characteristics protection, measurement control and performance test) grid standard, or a problem. It has been explained above: in January 2019, state Grid proposed the requirements of alternative meter positioning and expanded function design for "dual-core" smart meters, and "dual-core" smart meters will serve as the technology reserve of the new generation of smart meters. Therefore, the "dual-core" smart meter project, only four years before the implementation.

- -The starting design technology of the 2009 version of state Grid smart meter standard is not in place or incorrect positioning

First, smart meters, only give the definition, "smart" technology is not in place, how to implement the advanced interactive functions between the distribution network and users, it is not clear

· Internationally, there is no unified definition of smart meter. But some application features of the "smart" feature, some have in common:

A. In 2009, GE (China) published a document stating that GE smart meters have two bidirectional communication features and standard-based, open built-in advanced smart programs. As long as the information received conforms to the pre-set logic, it can make a judgment and response independently, without waiting for the main station to issue the command again.

B, 2009, Hangzhou haixing company published manuscripts points out: smart meter as a gateway communication application: uplink data communication, local data communication, and water meter, gas meter data communication, and home intelligent display unit (IHD) or customer information interaction unit (CIU) data communication, and the home second circuit relay control equipment communication.

C. In 2013, Germany formulated a new technical specification for the smart meter measurement system, and its smart core is the smart meter gateway. In this system architecture, the meter (and water meter and gas meter) is only the data measurement sensor responsible for the collection, while the smart meter gateway is the core of intelligence, which is the key functional module connecting the participants from different regions.

· The 2009 version of State Grid standard smart meter only has primary intelligent functions, mainly two-way power metering and grid measurement automation functions, basic power information interaction function between distribution network and users; the meter has no bidirectional communication mode, communication gateway application, and no advanced interaction function between distribution network and users. State Grid marketing / metering department and power distribution department have always had different views on whether smart meters directly support advanced interactive functions.

· The problems discussed here are directly related to the positioning of the state Grid electricity consumption information collection system. Internationally, AMI is an advanced measurement infrastructure, which aims to realize two-way, interactive communication and control between user side and power grid dispatching side, and the initial positioning of state grid electricity information acquisition system is to construct a remote automatic meter reading system of provincial power grid by transmitting electricity information collected from smart meters, which is essentially AMR system.

Second, the metering function positioning of the smart meter is not correct. The power metering function is not complete and solidified, which limits the requirements of other professional and provincial power grids according to the needs.

A. Internationally, single-phase and three-phase smart meters / electronic meters have single power and full power, which are selected by power supply departments or power users according to power trade settlement, grid line loss calculation, reactive power monitoring and control, and monitoring and control of harmonic pollution sources of users. The full power metering includes active power, reactive power, apparent power, and GE's kV2c three-phase smart meter design has distortion power factor calculation function. From 2021, Canada's electricity trade settlement will be measured by fundamental wave active power.

B. 2009 edition of state Grid standard smart meter: the single-phase smart meter only has active power metering function, and the three-phase meter is designed with active power and reactive power metering functions to meet the active power energy metering required by the current electricity price, and the maximum demand and power factor of three-phase users are calculated. At the same time, it is clear that the above measurement function cannot be changed.

State Grid, the 2013 version of the smart meter standard, the "dual-core" smart meter design scheme, and the technical requirements of the new generation of smart meter (paragraph 2), are all such unified metering function limits.

Third, the accuracy of single-phase smart meter is only level 2, and the factory error requires to be controlled within ± 0.6%, which is very unreasonable.

A. Internationally, the accuracy level of single-phase smart meter / single-phase electronic meter is designed to be level 1,0.5 and 0.2, which are selected as needed.

B. The author believes that the anti-power theft measures of low-voltage power grid should start from the single-phase smart meter measurement of 1kWh:

Users with a monthly electricity consumption of 500kWh or above need to install a level 0.2 single-phase smart meter;

Monthly electricity consumption of 200kWh or above users, install 0.5 level;

For users with monthly electricity consumption of 100kWh or above, install level 1;

Users with a monthly electricity consumption of less than 100kWh shall use level 2 or level 1.

At the same time, for users with a monthly electricity consumption of 200kWh or above, single-phase smart meters should be designed with a typical daily electricity consumption load curve. When the difference between the actual daily electricity consumption change of the user and the typical daily electricity consumption load curve exceeds the set limit, the meter shall send an alarm for future reference.

- -State Grid new generation smart meter (paragraph 2), it has been pointed out above: the design of this smart meter covers the requirements of the implementation of IR 46 standard, and in January 2019, State Grid proposed the design requirements of alternative meter positioning and expansion of functions. In January 2020, state Grid adjusted its development direction, and the positioning and expansion function design of the new generation of smart meters (the second paragraph) need to be re-approved and selected.

- -Improved local communication technologies supporting with the development of state Grid smart meters, including narrowband (low speed), narrowband (fast), broadband (medium frequency) power line carrier communication technology, micro-power wireless communication mode, two-way communication and gateway application and other technical issues. The author will write a special discussion.

**3. Expectations for the new generation of state Grid smart meters**

**1) The new generation of smart meters of state Grid needs to be finalized, improve the design, and then enter the large-scale trial operation for a long time in different environments**

-Reframe the design, including:

· The alternative meter positioning and expanded function design of the new generation of smart meters (paragraph 2) should be re-approved and selected;

· Supplement the new demand for the new generation of smart meters. As mentioned above, the key points of smart electricity consumption in 2020 proposed a new application of the new generation of smart meters, namely, research and development of smart Internet of Things home appliances, collection and transmission of smart energy information for households, and interaction and control of home appliances and the power grid. For these new demands, through function demonstration and verification, as the expansion function of the new generation of smart meters.

· Is it of practical value to apply the IR 46 standard in the active power metering of smart meters? How difficult is the comprehensive application of the IR 46 standard? Careful considered and re-evaluate.

-Expand argument, improve design: in view of a new generation of smart meters (second) completely changed the early smart meter integration architecture design, also new put forward some future application of metering function and many extension function (including meter software test), to ensure the quality of the new generation of smart meters and long-term application reliability, its metering department to organize meter full performance and function of expanded argument, improve design work, provide the basis for making new meter standard. On this basis, and then enter the scale of trial operation in different environments. It takes a long time, but it is safer.

- -up standard formulation and development of new technology topics

According to the related projects proposed by the author on November 19,2019, published by this article: State Grid: Multi-core modular Three-phase Intelligent Power Meter to Realize a New Leap in smart Meter Design Technology, the following development projects of the new generation of smart meter series enterprise standards and new technology projects are combined and refined:

· Type, function, technical requirements, testing and safety certification series standards of the new generation of smart meters designed with multi-core modular architecture;

· Algorithm, measurement traceability and application technical specifications of fundamental wave active measurement and harmonic active measurement;

· Technical specifications for orderly charging and metering of electric vehicles;

· Residential power load identification algorithm, assessment and application technical specification;

· Test requirements and application technical specifications for the new generation of smart meter software;

· Reliability technology demonstration and test of electricity meter service life of 15 years;

· DL / T698.45 Data communication protocol is applied to the revision of the new generation of smart meters.

**2) The re-development of state grid smart meters needs to make a long-term plan**

Reviewing from 2009 to 2020, the development of state Grid smart meters has gone through a tortuous development process and achieved many important achievements in the development and application of smart meters, but also left many problems requiring further research and forward-looking measurement.

- -From 2009 to 2020, the initial design defects of the development of smart meters need to be studied and repaired

· The "intelligent" functional technology of smart meters should be supplemented in place. Two-way communication and gateway technology are more widely used; the meter has the built-in advanced intelligent program for independent judgment and response to realize the advanced interaction between distribution network and users, including the interaction and intelligent control of household appliances of industrial and commercial users, industrial and commercial households, user demand response and future real-time price metering for real-time trading in power spot market.

· smart meter shall be open to the full power (active power, reactive power, apparent power) electric power metering function design, which shall be selected by other professional and provincial power grids according to their needs.

· Single-phase smart meters should be designed with multiple accuracy levels. Single-phase users shall evaluate the maximum monthly electricity consumption, and choose the accuracy level of the single-phase meter.

· Deepen the research on the smart meter software test technology, expand the test scope, improve the test efficiency, and formulate the technical specifications for the meter software test.

· Greatly reduce the replacement rate of smart meters at the initial stage of installation and operation, and study and improve product reliability design, test and demonstration.

- -Domestic high-end electricity meters have not fully entered the status of main meters at the grid threshold.

State Grid, the most influential central enterprise in China, has the most say in the operation quality of electricity meters at the import threshold of the power grid. The metering department of the State grid shall coordinate with the competent department of electricity meter industry to organize and coordinate the collaborative problems of domestic high-end electricity meters, mainly formulate the metering technical specifications of high reliability, high stability and long life of domestic metering meters at the pass; promote the design and production of domestic high reliability components; conduct the demonstration test of meter reliability design and the estimated operation of domestic high-end electricity meters and imported brand electricity meters on the Internet, accumulate long-term operation data, improve the design of domestic high-end electricity meters, and strive for the domestic high-end electricity meters to enter the status of the power grid pass.

- -Carry out forward-looking research on electric energy metering technology, accelerate the process of introducing new international metering technology, and improve the quality level of state Grid smart meters

, Its metering department jointly with the meter enterprise on the Internet and laboratory: at the same time to carry out the sine wave full power energy metering, very low power factor measurement, base wave measurement, nonlinear load measurement, direct 10-35kV high voltage metering advanced application technology research, its purpose is to improve the quality of its smart meter operation, reduce the grid line loss, inhibit grid harmonic pollution, improve large user measurement mechanism, provide measurement data.

Suggestion: measurement standardization administrative department of the meter industry international meter standards and communication and cooperation mechanism reform, its metering department, meter enterprises can have more opportunities to declare or directly involved in IEC, IEEE meter measurement standards and communication project, for meter industry in the international meter standard formulation and communication, at the same time, also conducive to its to speed up the process of the introduction of international measuring new technology.

· At present, the highest level and new trend of the development of smart meter and communication technology in the world

A. IEC is developing the standard for 0.1S static active energy meter; Langier is also developing 0.1S and E860 series.

B. The full power measurement accuracy exceeds the highest requirements of IEC standard: the active power measurement error is $\leq \pm$ 0.1%, the reactive power measurement error is $\leq \pm$ 0.2%, and the dependent power measurement error is $\leq \pm$ 0.3%.

C, internationally, the application of 0.5S class base wave reactive power measurement is being promoted. Canada: From 2021, electricity trade will settle electric energy, using base wave active power measurement mode.

D. Power quality monitoring technology will become the key point for the differentiated design and expanded application of the functions of electricity meters in the future.

E. Table measurement rate and communication: accurate and rapid measurement of instantaneous value, high-speed recording and high-speed wave recording; using multi-communication mode and multi-communication protocol application.

F, practical measurement and calculation technology of transformer and power line loss.

Application of g and Wi- - -SUN wireless communication technology: In March 2019, Langir Company won the bid for the deployment of more than 20 million smart meter projects of Tokyo Electric Power Company (Japan), connecting with user equipment through multi-communication technology and Wi- - -SUN home energy management communication standards.

-Its current a new generation of smart meter performance and communication technology design level, compared with the above international smart meter and new communication technology, there are many gaps, its metering department to run on the basis of the smart meter quality management experience, according to its energy new requirements for the construction of the Internet enterprise, the new technology of international measurement and communication, make its smart meter development plan.

**4. Domestic market demand and questions raised for the new generation of smart meters in 2021**

In the beginning of 2022, this paper summarizes 151 manuscripts and information about the development of the new generation of domestic smart power meter market and technology around the construction of a new power system. After combing, extract to promote the development of meter industry has important influence of 5 classes (domestic meter market, international meter market, new energy grid power quality monitoring and the metering of electricity, electricity information acquisition system technology update, promote the user side power Internet application) manuscripts is outlined, and put forward the subsequent need to discuss.

**1) In 2021, state Grid: The total bidding amount of new-generation smart meters increased by 28.2% year-on-year**

　　- -In 2021, the new generation smart meters of State Grid were invited for bidding in two batches (30.402 million in the first batch and 36.334 million in the second batch), and the total bidding amount was 66.736 million (the total winning amount was 16.97 billion yuan), an increase of 28.2% over the total bidding amount of smart meters in 2020 (52.056 million).

　　- -After summary analysis, the growth of the total bidding amount of the new generation of smart meters in 2021:

　　　The year 2021 is the first year of the application of the new generation of smart meters in the grid. 430 million smart meters (2009 and 2013) that do not meet IR 46 standards and the requirements of new energy grid metering need to be replaced. In an eight-year replacement period, an average of 53.7 million new-generation smart meters are needed per year.

　　　For the annual new users, based on 3% of the total number of users (510 million households) in the state grid power supply business area in 2020,15 million new tables are needed.

　　　New development of intelligent IOT electricity meter function batch application.

　　　Expand the application of the high-accuracy settlement pass table of the domestic power grid.

　　　For metering fault processing meters, calculated at 0.5% of the total number of electricity meters in operation of the State grid in 2020,2.55 million new meters are needed.

　　- -Here, for the first time, this paper focuses on the rising price of electricity meters. It is estimated that in 2021, the average unit price of the new generation of smart meters in China is 254.28 yuan, 19.6% higher than the average unit price of smart meters in 2020 (212.6 yuan).

　　- -Looking forward to the market prospects of the domestic new generation of smart meters in 2022.

　　As the market of the new generation of smart meter enters steady state and gradual period, in 2022, there are 376 million smart meters in operation (2009 edition, 2013 version), annual domestic demand update 53.7 million; including new market demand, new functions of smart meter electricity, it is expected that the minimum demand of the new generation of smart meters is 60 million in 2022. It can be said that the state grid electricity meter market is eternal.

　　However, state Grid has a limit on the maximum amount of electricity meter enterprises (about 6% of the total bidding amount), which is far from meeting the needs of sustainable development of electricity meter enterprises. Therefore, in recent years, on the basis of continuing to develop centralized bidding for electricity meters, and exploring the development of diversified industries. In 2020, smart meter exports grows rapidly, the marketing experience can be used, and domestic smart meters can enter the international high-end electricity meter market in batches.

**2) Promote the localization process of power grid gateway meters effectively**

　　In the past 20 years, Weisheng has successively introduced new high-end metering technologies such as 0.1S class active power measurement, nonlinear load measurement and harmonic measurement in the field of power metering.

　　In 2010, Weisheng developed the first 0.1S class with digital multiplier technology, with specific functions such as base wave, harmonic measurement, satellite timing and tide crossing. It is compatible with various communication protocols and supports multiple language display.

　　The pass table adopts independent intellectual property measurement algorithm (including multiplexing Newton- -Cotes high order integral algorithm, reactive power measurement algorithm, high precision dynamic angular difference compensation algorithm), selects international high-performance components, and uses advanced production process to ensure high accuracy of high-end measurement products (active power measurement factory error less than / equal to + / -0.04%; reactive power measurement 0.5S class), high stability (active power measurement error less than / equal to + / -0.1% in 15-year life cycle) and high reliability. After testing, compared with the performance of similar international power grid pass table, the domestic 0.1S class pass table has reached the international advanced level, breaking the monopoly position of the import pass table, and promoting the localization process of high accuracy settlement pass table.

At present, Weisheng 0.1S class pass table has been used for trans-regional power grid connection line hub, inter-provincial power grid connection line substation, large capacity new energy power generation online measurement, impact load electrified railway traction station and other pass measurement. It has been running for many years, and the measurement performance is stable and reliable.

**3) The intelligent iot electricity meter with non-interventional load identification function is applied in the first batch of Hunan power grid**

In July 2021, Hunan Electric Power Company installed 543 smart IOT electricity meters in the first batch of five low-voltage platform areas in Changsha and Xiangtan. The table is equipped with a non-interventional load identification module, which can obtain the high-frequency sampling waveform of voltage and current. By using the electrical feature combination algorithm, the user's electrical power consumption and electrical consumption period can be identified.

On the basis of the traditional smart meters, the smart IOT electricity meter adds the functions of harmonic measurement, new temperature measurement and measurement error self-monitoring, and its service life is increased from 10 years to 16 years. Therefore, the meter can predict the hidden danger of heating and burning, solve the metering blind area of high-speed railway and electric power supply, and high-harmonic pollution equipment connected to the power grid, so as to prevent various risks.

In 2021, Hunan Electric Power will promote the application of more than 1,000 smart IoT electricity meters in the province.

**4) Jiangsu Electric Power: Launch a multi-core modular load analysis and electricity meter that can be remotely upgraded software**

-In the previous non-interventional load analysis function, under the framework of the existing electricity information collection system, the new function is upgraded to replace the load analysis module on the site, which is difficult to meet the requirements of large-scale application.

- -In the existing domestic standard system, online software upgrade is not allowed.

- -However, according to the IR 46 standard, only when the independence of the metering core and non-metering core can we ensure that the function expansion of the non-metering part and remote upgrade do not affect the measurement performance, which provides ideas for the development of load analysis power meters for remote upgrade software.

- -The new electricity meter is designed with metering core, universal function core and special load analysis core. Through the interaction of electricity information collection system, the normal metering core function and the online update of household load identification function

In the new remote system, the communication topology; the remote upgrade process of load matching algorithm and load feature library.

- -Power meter load analysis chip, using a multi-layer tree classification algorithm based on Mahalanobis distance.

-After test, the identification accuracy of single appliance can reach 95-98%, more than 70-90%, indicating that under the premise of not affecting the measurement performance, the load analysis software for remote upgrade can accurately identify the details of typical household appliances.

**5. Internationally, there is still much room for development in the demand for smart meters, and the competition in the international electricity meter market is more fierce, and the overall export of domestic high-quality smart meters and AMI / AMR system is promoted**

Description: AMI (Advanced Metering Infrastructure) Advanced Measurement Infrastructure.

The AMR automatic meter reading system.

**1) International electricity meter market summary**

-The total population of 230 countries is about 7.5 billion; 89% (6.675 billion) and the total number of households is about 1.335 billion.

- -At present, there are about 1.7 billion electricity meters in use, including industrial, commercial and residential electricity meters; in 2020, about 963 million smart meters are installed and applied, and the penetration rate of smart meters is 59%, that is, the global demand for smart meters is 737 million, and the overall international smart meter market maintains a continuous growth trend.

- -Internationally, the smart meter market started in Europe and the United States. After 10 years of development, China currently accounts for 62% of the global smart meters, followed by North America, Europe, Asia (excluding China), Latin America, the Middle East and Africa.

- -Internationally, at present, smart meter companies with a good reputation and a large export volume:

Langier Corporation (Landis & Gyr)

Alley (Itron)

General Electric Corporation (GE)

Elster Corporation (Elster) et al

- -domestic smart meters

In 2020, Chinese electricity meter enterprises exported electricity meters to 160 countries and regions, exporting about 59 million single and three-phase smart meters, the total export amount of more than 1.3 billion US dollars (8.4 billion yuan), and the average export price of electricity meters was 30.7 US dollars (200 yuan).

Chinese electricity meter enterprises with large exports of domestic smart meters: Hangzhou Haixing, Weisheng, Shenzhen Technology, Huali, etc. Among them, Haixing exported 280 USD million (1.8 billion RMB) in 2020.

**2) Internationally, the century-old electricity meter brand: Langier high-quality smart meter**

- -In 1993, Langier produced the ZU-type 0.2S class high-precision pass table, using the time-split multiplier.

- -In 1995, Langier introduced the ZB type 0.5S class stationary three-phase table based on the Hall multiplier (DFS).

- -In 1998, Shandong Power Grid imported the power grid electricity automatic charging system produced by Langier. The power plant has ZU 0.2S pass meter and FAT data processor; the high voltage substation, ZB 0.5S static electricity meter and METCOM data processor.

- -In 2004, Langier developed a new generation of pass metering quality- -ZQ type 0.2S class high-precision settlement pass table, using digital multiplier technology.

- -In the early part of the century, the AMI (Advanced Measurement Infrastructure) and its command system (Grid Stream) were developed.

— In 2016, Langier developed a grid router; a grid router-based ability to connect to multiple grid deployment AMI solution (Grid Stream) communication platforms.

- -In 2019, TEPCO will cooperate with Langier to complete the AMI project in 2020, installing and applying more than 20 million smart meters, using the Wi-SUN mesh network communication protocol that meets the IEEE802.15.4 standard

-In 2021

On September 28,2021, Rangel Group announced that its subsidiary Landis & Gyr AG has reached a merger agreement with (Turkey) Luna Company- -Rangel will acquire% of Luna100 shares, and the acquisition is expected to complete by the end of 2021.

Luna is a leading meter supplier in Turkey and has established a dominant position in engineering, certification, and manufacturing of energy metering, with annual revenue of over $60 million.

The acquisition of Luna will give Langir a significant share of the Turkish market, including AMI business opportunities that are not yet launched, as well as using Luna's vertical integration capabilities to boost sales in the surrounding areas and markets. Currently, Luna products are exported to many countries in Africa, Asia, Europe (including Germany) and South America.

On July 20,2021, "Transmission and Distribution World" reported: Landis & Gyr, an American subsidiary of New Jersey Public Utilities (PSE & G) and Landis & Gyr Group (Langier Group) AG

Technology lne Signed a 10-year agreement. Langier will provide 2.3 million smart meters to PSE & G, as well as related network infrastructure, software & services. All the meters are connected to the Langier Grid stream Connect AMI platform. PSE & G will complete the deployment of 2.3 million smart meters over the next four years.

The Langir Grid stream network is also used to support sensors for grid management and consumer interaction technology for power and gas services; Grid stream AMI adds grid edge intelligence to every metering and network device to improve efficiency and automation for utilities and consumers.

- -Focus on Langier is transforming

In April 2017, suge reported by Switzerland: Langier will transform from a professional metering supplier to a company providing customers with the industry's most advanced network and Internet of Things (IOT) solutions.

Langier, annual sales of $1.6 billion, is (Japan) Toshiba's independent development platform, by the Japanese innovation network company (INCJ) holds a 40% stake, is committed to provide comprehensive solutions for the basis of the smart grid, including smart meters, distribution network sensors and automation tools, power load control, analysis and power reserves, in 5 continents more than 30 countries do business, promote the global better management of energy.

**3) Promote the overall export of domestic energy and high-quality smart meters and AMI / AMR for the needs of global energy and new energy management**

First, in April 2021, the Saudi smart meter project of State Grid was officially completed, realizing the first large-scale overseas export of China's electricity information collection system business

- -Summary of the Saudi smart meter project

The general contractor of the project is China Power Technology and Equipment Corporation of State Grid (hereinafter referred to as CLP Equipment Corporation)

The project was signed on December 19,2019, involving nine regions in the western and southern part of Saudi Arabia, including the main station system, 5 million smart meters, concentrators, external circuit breakers, supporting software development and hardware equipment scheme design, production, supply, transportation, storage, installation, commissioning, testing, etc.

Previously, in April 2019, Nanrui Group started the key technical breakthrough of the technical scheme verification of Saudi smart meter project, and organized the testing work of AMI / AMR integration and interconnection of AMR system.

In July 2019, Nanrui Group led the lead to provide the overall AMI / AMR solution, which stood out in the bidding organized by Saudi Power Company, laying the foundation for CLP Equipment Company to win the winning project.

The project was purchased in batches in China, and Linyang won RMB 330 million; Ningbo Samsung won RMB 450 million; Chint supplied shell plastic circuit breaker over RMB 100 million; Shenzhen completed 2.4 million smart meters within 1 year; Nanrui became the contractor of power information collection system of the project and was responsible for promoting the implementation of the project.

- -Nanrui, develops and outputs a set of AMI / AMR system whole-process business system according to local conditions, and puts forward an overall solution suitable for local special needs

In March 2020, the owners requested the addition of narrowband Internet of Things (NB-IOT) communication mode, and the application ratio increased from 5% to 50%. The number of NB-IOT access equipment at the same time has a great impact on the communication quality. After joint research, the smart meter reading scheme suitable for millions of NB-IOT was finally designed.

The project is based on the main station system architecture such as intelligent acquisition elastic architecture, massive data storage and intelligent data analysis, which can meet the communication connection management of 7.5 million smart meters, heterogeneous communication access, and 1 billion metering data storage and processing requirements.

The terminal information communication of the project adopts the dual-mode and narrowband wireless communication scheme.

- -Shenzhen Technology's Chengdu Great Wall Development Company exported 2.58 million smart meters (15 years of service life) to the Saudi smart meter project, accounting for the largest share of the smart meters in the project.

In the past two years, Middle Eastern countries have a large demand for energy transformation, and have begun to deploy smart meters on a large scale, requiring advanced communication technology, dual-mode operation, suitable for local environmental applications.

On the one hand, the company directly exports finished electricity meters and remote control equipment through the processing trade outside Chengdu; on the other hand, after exporting the semi-finished products, assembling the smart factory with the company as the starting point to build the local intelligent production system and expand the export market.

Second, Weisheng: from "product export" to "overall solution output"

- -In August 2021, Weisheng Brazil signed the smart meter winning contract with the Brazilian Electric Power Authority, with a total amount of more than 100 million yuan and the number of more than one million meters.

The bid reflects the internationalization of Weisheng brand and international customer recognition.

- -In September 2021, Weisheng Mexico signed a contract with the Mexican Federal Electricity Commission to supply more than 1 million smart meters.

The total amount of this contract is over 260 million yuan, and the product delivery work will be completed in the next few months

Weisheng, the recent strategy in the global market has achieved considerable results, which is based on the results of the sustainable market

- -Weisheng is gradually becoming a global supplier of overall solutions in the energy Internet of Things and management field

In 2020, the total operating amount of Weisheng was 4 billion yuan, up 8% year on year. Overseas business accounts for 15% of the total business.

Weisheng, international for 20 years. At present, the product field has expanded from power metering equipment to water, gas, heat and electrical products; the business field has expanded from energy metering to intelligent management; and the business model has expanded from product management to overall solutions and engineering services.

Since 2014, Weisheng has been working in overseas markets for many years, familiar with the national conditions and market operation rules of countries along "Belt and Road", and can better grasp the investment direction; quickly open marketing channels through "localization" operation. Currently, the products have been sold to more than 50 countries and regions; factories in Tanzania, Brazil and Mexico.

Third, the comments on this article

Title: It is the right time to promote the overall export of domestic high-quality smart meters and AMI / AMR!

As already described before, the domestic, state grid electricity meter market is eternal. However, at present, the state grid electricity meter market has entered a stable and gradual period, and the demand for smart meters can no longer meet the needs of the sustainable development of electricity meter enterprises.

In the past 10 years, electricity meter enterprises have made great achievements in exploring the export of smart meters. At present, China's electricity, water and gas meter exports account for 50% of the global market share. The export method mainly includes product export; semi-finished products are exported and then assembled in overseas area; overseas factories or joint ventures, mainly in Asia, Africa and Africa.

This time, the Saudi smart meter project of State Grid is a typical case of the overall export of smart meter and electricity information collection system (including the main station system), which is the first time in China.

- -The United States, the development of AMI and the questions raised

This part is extracted from "Introduction to Smart Grid" edited by Xu Xiaohui and edited.

In August 2006, the Federal Energy Policy Commission defined the concept of AMI.

AMI mainly includes home network system, smart meter, local communication network, communication network connecting to the data center of the power company, meter data management system and data integration platform, etc.

AMI is a fully configurable infrastructure and integrated into current and future grid and operational processes.

Smart meter: it is a kind of green meter, it has the demand response function, can reduce $CO_2$ emissions, save energy, improve energy efficiency.

Smart meter, programmable metering device, its main functions: time-sharing electricity price; providing power consumption for users and power companies; net metering price for installing renewable energy generation devices; notification of power failure and power outage recovery; remote on and off functions; prevent high electricity price or peak load limit for demand response service; pre-payment; power quality monitoring; anti-theft monitoring; and other intelligent equipment communication.

Communication facilities: to provide a continuous information interaction function between power companies, users, and controllable loads. It must be an open two-way communication standard and highly reliable.

AMI must have a local information concentrator to collect data from all smart meters and transmit these data through the channel to the central server. The media used for communication can be diversified.

The downlink of the "last mile" problem from concentrator to electricity meter is a technical bottleneck. There is no feasible solution, mainly because the "last mile" faces a series of problems, such as large quantity, complex field environment, low quality of communication media, and high cost pressure.

• home-area network

In AMI, the main functions of household energy management are: real-time display of energy usage and price information; feedback of electricity price information; setting energy consumption or load control peak; automatic load adjustment; user human control function.

The main communication methods suitable for meter and home equipment: BPL technology, Home Plug technology, Zig Bee technology, and WiFi, WiMax technology are also home network technology.

In AMI, the home network interacts with the user portal (it can be installed in any device), connects the smart meter and controllable electrical equipment, and combining the home energy consumption information of the power company.

Meter Data Management System (MDMS): It is a database with analysis function that can interact with other information systems. Its main function is to verify, edit and evaluate the legitimacy of AMI data.

Operation gateway: AMI obtains the information required for AMI functions by interacting with the application layer of many systems of the grid (advanced distribution system, advanced transmission operation system, advanced asset management system).

How does AMI support the smart grid operation requirements?

First, enhance the initiative and enthusiasm of users to participate in the power grid.

Second, real-time monitoring and control of distributed power generation and energy storage equipment around users.

Third, contact the users and the power grid to increase the activity of the market. Users should take the initiative to participate in the power grid, adjust the load or transfer the energy to the power grid according to the price information.

Fourth, the smart meter is equipped with power quality monitoring module, which can quickly measure, diagnose and adjust power quality.

Fifth, the distributed power grid operation model can be realized to reduce the external impact on the power grid attacks.

Sixth, to assist the power grid to achieve fault self-healing.

Seventh, provide accurate and timely data information to better improve power grid asset management and power grid operation.

- -Rangel AMI development process has been described previously; on December 19,2012, Rangel (Zhuhai) Company organized the Rangel AMI technology exchange meeting.

At the meeting, rangier's global AMI pilot situation was highlighted.

The AMI pilot project in the United States: using RF Mesh wireless network technology.

The AMI pilot project in Europe: the adoption of the Prime-G3 narrowband (fast) power line carrier network technology.

AMI software system: Grid stream.

- -Requirements for the overall development of domestic high-quality smart meters and AMI / AMR:

Development and export of domestic high-quality smart meters

At present, the price of smart meters in China is relatively low. With various measures, it can reach the international medium quality level, but it has not entered the international high-end electricity meter market in batches.

There is no unified standard for high quality smart meters. For a long time, after the performance test of imported smart meters and pass meters, the index requirements of imported high-quality smart meters can be summarized:

First, the factory error of smart meter is strictly controlled, generally 40% of the meter grade; 25%.

Second, the impact volume index, control in 50% of the standard requirements.

Third, the meter error curve is flat, and the linearity is 20% of the meter grade.

Fourth, the style is flat, generous, and the surface temperature rise is low.

Fifth, the service life period of the meter, to ensure that the error does not exceed the grade requirements.

Sixth, Langier pass meter: the use of independent development of electric energy metering algorithm proprietary chip, to improve the reliability and safety of the meter is very effective.

Design of the AMI instruction system and the master station / platform. The main station system of Saudi Arabia smart meter project is based on 10 years of experience in the main station design of domestic electricity information acquisition system. However, most domestic electricity meter enterprises do not have the experience in this field, which is a short board. In the future, they need to increase the investment in development and the gathering of high-level software talents to make up for this lesson.

AMI instruction system, main station / platform design, can refer to the design experience of Langier and Nanrui, collect the AMI function requirements of different regions in the world, with the flexibility that the system can be modified and supplemented.

User Internet of Things application design, international and domestic is in the beginning, the development process. Among them, the local communication adopts HPLC, Wi-SUN, narrow band wireless communication technology, which needs to adapt to the residential density of local residents, climate environment requirements, communication media, etc., and puts forward the innovative design of user Internet of Things application. Among them, the Wi-SUN communication protocol is expected to become the core protocol of the Internet of Things. As mentioned earlier, the 20 million smart meter project deployed by TEPCO and Langere uses an Internet of Things network that is compatible with smart devices through the use of various communication technologies and Wi-SUN home energy management interoperability standards. Due to the end-to-end addressable capability, wide coverage, security, and high scalability, the Wi-SUN will expand to smart homes.

**Conclusion**

In the near future, with the continuous progress of science and technology, our life will become more convenient and efficient. Among them, the smart meters, as an important part of the smart grid, are also constantly evolving and upgrading.

With the continuous development of technology, the function of smart meters has also been continuously expanded. Modern intelligent electricity meter not only has the basic functions of electricity metering and remote meter reading, but also can realize load control, user information management, power quality detection and other functions. Through smart meters, grid operators can have a more comprehensive understanding of their electricity consumption and provide users with more personalized services.

At the same time, with the continuous development of the Internet of Things, cloud computing, big data and other technologies, smart meters have also begun to combine with these technologies, forming a more intelligent and efficient power grid management system. Through the electricity consumption data collected by the smart meters, the power grid operators can analyze, predict and optimize the electricity consumption to improve the power supply efficiency and reliability of the power grid.

There are also some important technological trends to watch in the evolution of smart meter technology. The first is intelligence and networking. In the future, smart meters will pay more attention to the interaction with users and intelligent decisions, so as to achieve more personalized and intelligent electricity service. The second is safety and reliability. With the increasingly prominent problem of network security, the security and reliability of smart meters have also become an important technical trend. Finally, green protection and sustainable development. smart meters will pay more attention to environmental protection and energy conservation, promote the transformation of energy utilization mode, and achieve sustainable development.

In short, with the continuous progress of technology and the broadening of the application field, smart meters will play a more important role in the smart grid, and make a greater contribution to the development of the power industry and the improvement of energy utilization efficiency. At the same time, we also need to pay attention to the safety and reliability of smart meter technology to ensure the safety of users' electricity and data. In the future, smart meter technology will continue to evolve and innovate, bringing us a more convenient, efficient and sustainable electricity experience.